\documentclass{article}
\usepackage{spconf,amsmath,graphicx,bm}
\usepackage{booktabs}
\usepackage{caption,subcaption}
\usepackage{etoolbox,siunitx}
\usepackage{booktabs}
\usepackage{multirow}
\usepackage{amssymb}

\title{Improving Source Separation by Explicitly \\ Modeling Dependencies Between Sources}

\name{Ethan Manilow$^{1,2,}$\sthanks{Work done as a Google Student Researcher.}, Curtis Hawthorne$^{1}$, Cheng-Zhi Anna Huang$^{1}$, Bryan Pardo$^{2}$, Jesse Engel$^{1}$}
\address{$^{1}$Google Research, Brain Team \quad $^{2}$Northwestern University}

\begin{document}
\ninept
\maketitle
\begin{abstract}
We propose a new method for training a supervised source separation system that aims to learn the interdependent relationships between all combinations of sources in a mixture. Rather than independently estimating each source from a mix, we reframe the source separation problem as an Orderless 
\underline{N}eural \underline{A}utoregressive \underline{D}ensity \underline{E}stimator (\textsc{Nade}),
and estimate each source from both the mix and a random subset of the other sources.  We adapt a standard source separation architecture, Demucs, with additional inputs for each individual source, in addition to the input mixture. We randomly mask these input sources during training so that the network learns the conditional dependencies between the sources. By pairing this training method with a block Gibbs sampling procedure at inference time, we demonstrate that the network can iteratively improve its separation performance by conditioning a source estimate on its earlier source estimates. Experiments on two source separation datasets show that training a Demucs model with an Orderless \textsc{Nade} approach and using Gibbs sampling (up to 512 steps) at inference time strongly outperforms a Demucs baseline that uses a standard regression loss and direct (one step) estimation of sources.

\end{abstract}
\section{Introduction}
\label{sec:intro}

Within an auditory scene, musical sources are highly coordinated; musicians create sounds that are intended to overlap in time (i.e., rhythm) and frequency content (i.e., harmony). This coordination is fundamental to most music and helps distinguish it from other types of auditory sources such as speech and environmental sounds. However, this coordination makes isolating individual sounds difficult, as done in music source separation. The synchronization of harmonic content can lead to overlapping frequency partials; if different instruments produce different notes in a chord voicing, it might be hard to determine which frequency belongs to which source. For example, the root of a chord might be played by a bass guitar and the rest of the chord played by a piano. In this case, the higher partials of the bass might be hard to distinguish from those that came from the piano.

It stands to reason, therefore, that when doing source separation complete or partial knowledge about one source could potentially be helpful when estimating another. This contextual knowledge might be essential in the case of the bass guitar and piano above: if all or some of the partials of the piano notes are known, a system might be able to make a better estimate of the bass. In probability theory, this process is termed ``explaining away'', where possibilities are eliminated as knowledge is acquired. While explaining away might be especially important for musical signals, we note that our approach might be broadly useful even in unstructured audio scenes.

Despite this, deep learning-based music source separation research has largely ignored complementary context between musical sources, opting instead to model each source independently. More often than not, this means a separation system is designed to separate a single source given only the mixture as input, with no consideration given to explaining away information provided by other sources. In this paper, we introduce a new means of training and sampling from a source separation system such that it is able to leverage contextual information provided by other sources.

\begin{figure}[t]
    \centering
    \includegraphics[width=0.95\columnwidth]{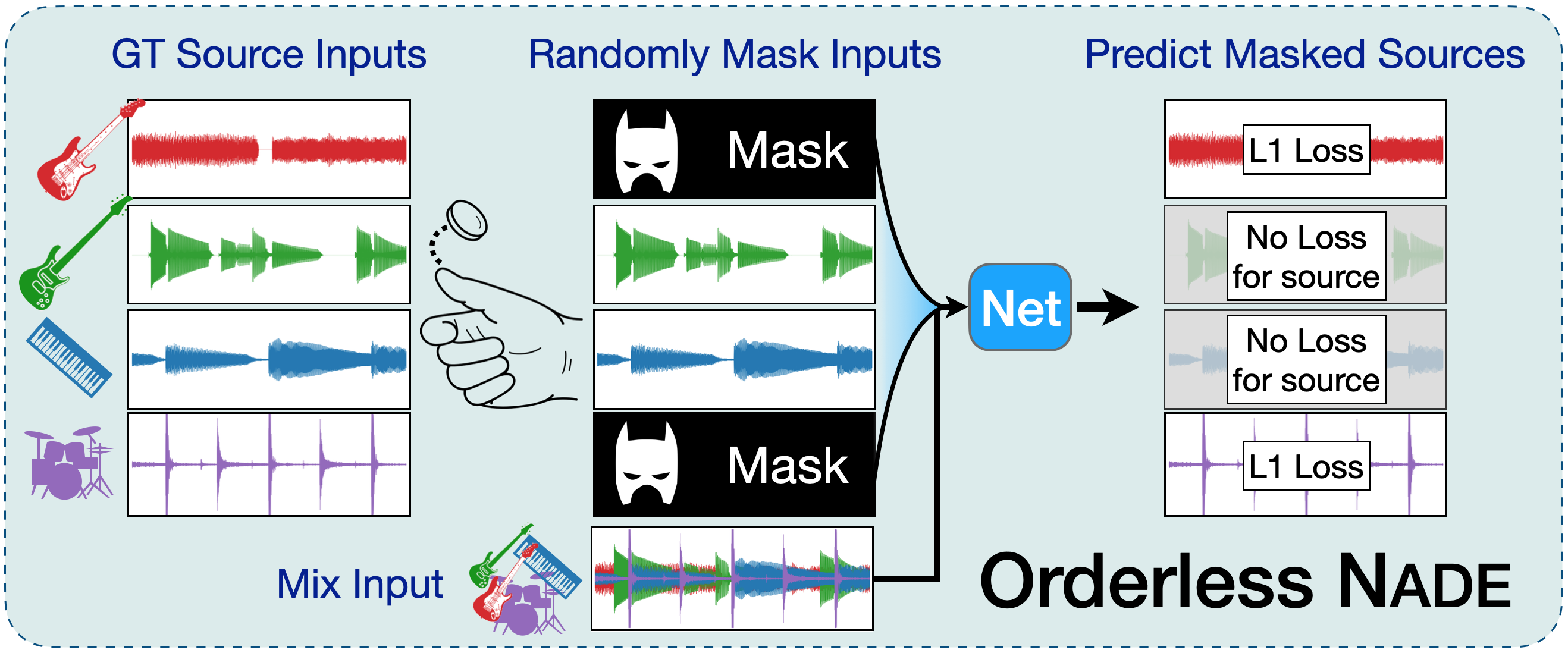} \\
    \vspace{0.1cm}
    \includegraphics[width=\columnwidth]{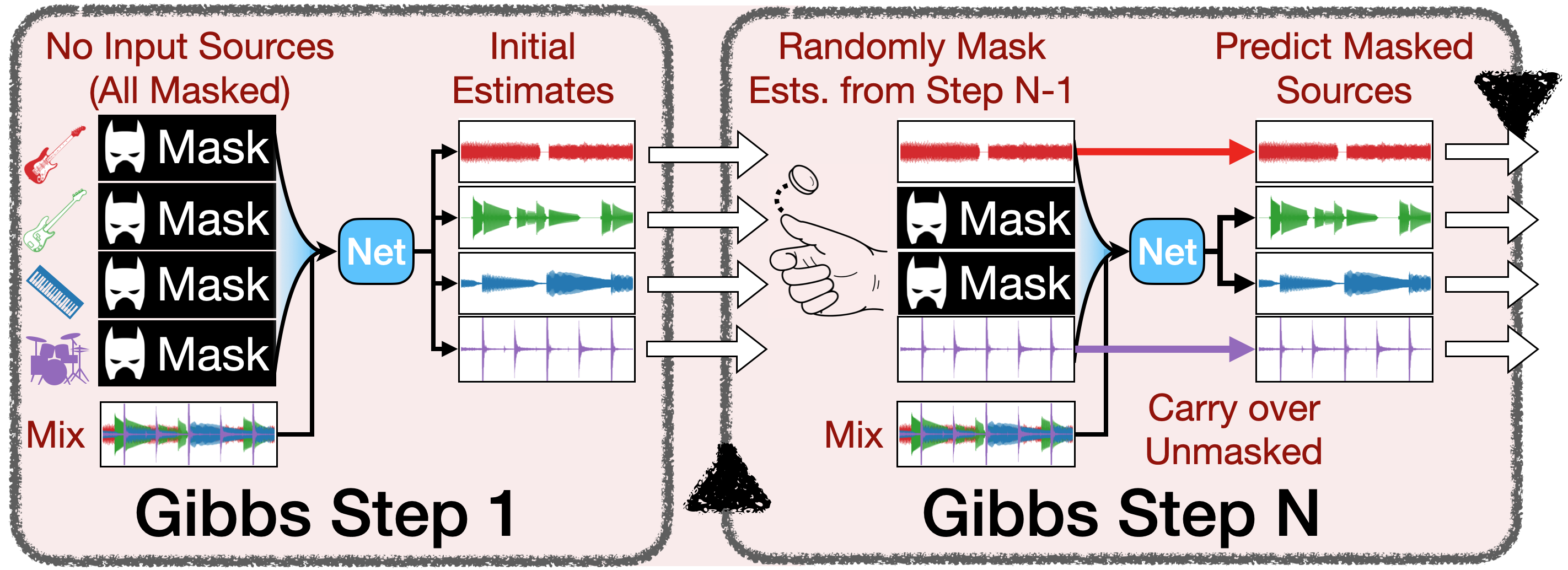}
    \caption{The proposed Orderless \textsc{Nade} training (top) and block Gibbs sampling procedure (bottom). For Gibbs Step N, un-masked source inputs are the model's source estimates from Step N-1.}
    \label{fig:diagram}
    \vspace{-0.5cm}
\end{figure}

We propose alterations to the training and inference procedures of existing neural networks to account for contextual information between different sources in a musical scene. Specifically, we propose to model the dependencies between different sources within a mixture by autoregressively factorizing the conditional probability distribution, $p(\mathbf{s}_1, \dots, \mathbf{s}_I | \mathbf{m} )$, of a mixture $\mathbf{m}$ and sources, $\mathbf{s}_1, \dots, \mathbf{s}_I$, in such a way that a model learns useful conditional dependencies between all combinations of sources within a mix. To accomplish this, we alter a separation network with additional waveform inputs used for conditioning the net with contextual source data. During training, we provide ground truth source data to a random subset of these conditioning inputs (i.e., teacher forcing) and task the network with estimating the sources not given as input following an Orderless \underline{N}eural \underline{A}utoregressive \underline{D}ensity \underline{E}stimator (\textsc{Nade})~\cite{larochelle2011neural,yao2014equivalence} formulation (see Section \ref{sec:onade}). At inference time, we run a block Gibbs sampling procedure by first making a set of initial source estimates, and then iteratively feeding those estimates back into the network to refine and improve them.

This training and inference procedure is agnostic to the choice of model architecture, therefore we experimentally verify our method using an existing music separation network. We show that the proposed training and sampling procedure substantially increases separation performance on two source separation datasets. We show the effect of Gibbs sampling and, in many cases, the performance of source estimates monotonically increase no matter how many Gibbs sampling steps we try. In other cases, source estimates improve over baseline training after one inference step (i.e., no Gibbs sampling), indicating that our Orderless \textsc{Nade} training is beneficial by itself.

\section{Prior Work}
\label{sec:prior_work}

In the typical source separation formulation, the goal is to predict some source given an input mixture. In other words, to produce a source estimate, a model is \textit{only} conditioned on the mix, making each estimate produced independently of any other source in the mix.

This formulation leads many systems to treat each source independently, leading some top performing systems to independently train a new network for each source (i.e., ``one-vs-all'')~\cite{mimilakis2018monaural,stoter2019open,hennequin2020spleeter}, effectively treating each network as a source denoiser. Other systems output multiple sources at once (i.e., ``multi-source'')~\cite{manilow2020simultaneous, sawata2021all}, potentially enabling a net to implicitly learn relationships between the sources, but ultimately still only conditioning the net on the mix. In this work, we explicitly condition a net on other sources in the mix. 

A number of recent music separation systems condition on auxiliary information as a means of assisting the separation process. For instance, some systems condition on the linguistic content of singing \cite{chandna2020content}, the musical score \cite{miron2017monaural}, class of the musical source~\cite{MeseguerBrocal2019, slizovskaia2019end, seetharaman2019class, samuel2020meta}, or an audio query input~\cite{lee2019audio, manilow2020hierarchical} as a means of steering the network to separate different sources. In this work, we condition on other sources within the same mixture as a means of learning dependencies between different sources in a mix, without auxiliary information. In Section~\ref{sec:gibbs} we describe how this enables a model to increase its separation performance by bootstrapping from its initial estimates. 

Some works have proposed iterative inference methods for source separation, similar to our work. For instance, Mimilakis et. al.~\cite{mimilakis2018monaural} propose a way to iteratively estimate a mask for singing voice separation, a one-vs-all separation setup. In the multi-channel case, iterative setups have been proposed for music separation by emulating the EM algorithm in order to create a multi-channel Wiener filter~\cite{nugraha2016multichannel} or by estimating a multichannel mixing matrix~\cite{mogami2018independent}. Our work is not dependent on the additional information provided by multi-channel mixtures (e.g., interchannel level differences); we assume all signals are mono. In the speech separation literature, iterative methods have been used to remove background noise from overlapping speakers before separating them~\cite{Wichern2019Interspeech09}, and also to separate multiple speakers one by one~\cite{takahashi2019recursive}. Here we focus on music separation. In general, our work can be seen as related to the iterative procedure proposed by Deep Equilibrium Models (DEQs)~\cite{bai2019deep, bai2020multiscale}, although our system is able to be optimized with traditional SGD-based methods (e.g., Adam~\cite{kingma2014adam}) and our iterative inference procedure is a result of our goal of trying to learn dependencies between sources in a mix, not try to find a fixed equilibrium points, which is the goal in DEQs.

Some recent work has taken a more generative modeling perspective on source separation, similar to our proposed autoregressive approach. For instance, Jayaram \& Thickstun~\cite{jararam2021parallel} propose a fast sampling method for a WaveNet~\cite{oord2016wavenet} such that it can be used in a one-vs-all separation setup. WaveNet is a causal, autoregressive generative model for raw audio that produces a waveform, $x$, such that the output sample $x_t$ at time $t$ is conditioned past predicted samples, $p_\theta(x_t |x_1, \dots, x_{t-1})$. This setup is autoregressive \textit{in time} (within a source) but not \textit{between} sources. In contrast, our work does not model dependencies in time (within a source), but does model dependencies between sources.
Finally, Demucs~\cite{defossez2019music} is a source separation network inspired by generative modeling, whereby the network forgoes a masking step in favor directly estimating the waveforms of multiple sources. In this paper, we extend Demucs v2 to model dependencies between sources.

Our work draws inspiration from \textsc{Coconet}~\cite{huang2019counterpoint}, which is a generative model of musical scores and uses contextual note information to infill new musical voices. We expand the scope of their Orderless \textsc{Nade}~\cite{larochelle2011neural,yao2014equivalence} formulation to the continuous domain of audio signals, by modeling sources in a source separation context. Our method can alternatively be viewed as similar to a masked language model (MLM) setup, used for training models like BERT~\cite{Devlin2019BERTPO} and T5~\cite{raffel2020exploring}. These models mask words to learn dependencies within language corpora, whereas we mask sources to learn dependencies within mixes. Our work can also be seen as fitting under the umbrella of order agnostic diffusion models~\cite{austin2021structured, hoogeboom2021autoregressive}, the main difference being that in our case each source is considered a discrete maskable variable, albeit with many continuous dimensions.

\section{Orderless NADE Training}
\label{sec:onade}

Source separation is the task of conditionally predicting a set of sources $\mathbf{s}_1, \dots, \mathbf{s}_I$, given a mixture, $\mathbf{m}$, like $p_\theta(\mathbf{s}_1, \dots, \mathbf{s}_I | \mathbf{m})$, for some model $p_\theta$ with parameters $\theta$. Typically, this conditional distribution is independently factorized (Eq.~\ref{eq:indep_factorization}), resulting in an independent training objective (Eq.~\ref{eq:indep_loss}):

\begin{equation}
    p_\theta(\mathbf{s}_1, \dots, \mathbf{s}_I | \mathbf{m}) = \prod_{i=1}^{I} p_\theta(\mathbf{s}_i | \mathbf{m})
    \label{eq:indep_factorization}
\end{equation}
\begin{equation}
    \mathcal{L}_{indep.} = - \sum_{i=1}^{I} \log p_\theta(\mathbf{s}_i | \mathbf{m}).
    \label{eq:indep_loss}
\end{equation}
For example, the L1 loss used by Demucs corresponds to estimating the negative log likelihood of the waveform under a Laplacian distribution.
Many systems even train one network per source, solidifying the independent factorization as a hard prior into the system's design. We note that the dimensionality of the sources $\mathbf{s}_1, \dots, \mathbf{s}_I$ and mixture $\mathbf{m}$ are intentionally omitted because this formulation is agnostic to their domain (i.e., waveform, spectrogram, etc).

However, in this work use a different way to factorize this conditional dependency,
aiming to explicitly capture coordination between sources. As such, we want to condition each source on all possible combinations of other sources in the mix. To accomplish this, we propose a new way of training source separation systems.

Just as before, we want to conditionally predict a set of sources given a mix, $p_\theta(\mathbf{s}_1, \dots, \mathbf{s}_I | \mathbf{m})$. Specifically, for the set of sources $S = \{\mathbf{s}_1, \dots, \mathbf{s}_I\}$, at each training iteration we randomly draw a subset of the sources, $D \subseteq S$, and then condition on this subset to predict sources in the complement set, $\lnot D$. With $\bf D$ denoting the set of all random subsets of $S$, this enables us to factorize the conditional probability distribution like

\begin{equation}
    p_\theta(\mathbf{s}_1, \dots, \mathbf{s}_I | \mathbf{m}) = \prod_{\bf D}^{} p_\theta(\mathbf{s}_{\lnot D} | \mathbf{s}_{D}, \mathbf{m}).
    \label{eq:onade_dist}
\end{equation}

As it turns out, this is an Orderless \underline{N}eural \underline{A}utoregressive \underline{D}ensity \underline{E}stimator (\textsc{Nade})~\cite{larochelle2011neural,yao2014equivalence}, where we autoregressively predict sources in $\lnot D$ conditioned on sources in $D$. As such, the loss is given by

\begin{equation}
    \mathcal{L}_{\text{O-\textsc{Nade}}} = \displaystyle \mathop{\mathbb{E}}_{D \sim \bf{D}} - \sum_{i \in \lnot D} \log p_\theta(\mathbf{s}_{i} | \mathbf{s}_{D}, \mathbf{m}),
    \label{eq:onade_loss}
\end{equation}

\noindent where we apply a loss as in Eq.~\ref{eq:indep_factorization} to those sources in the compliment set $\lnot D$. Note that this factorization requires that the model $p_\theta$ have different inputs than the factorization in Eq.~\ref{eq:indep_factorization}: whereas before, the model only needed the mixture as input, in Eq.~\ref{eq:onade_dist} the model now needs the mixture \textit{and} the subset of sources in $D$ as input. Because the sources in the complement set $\lnot D$ are not used as input conditioning, we say that these sources are \textit{masked}.

By masking a subset of its source input, the network must learn to use the unmasked part of its input as a conditioning signal for predicting the masked sections. With the proposed training setup the network learns the context surrounding the masked segments: it must use the unmasked context to predict the masked segments, therefore learning dependencies between sources.
During training, loss is only computed for the source estimates that are masked, and the additional source inputs are teacher forced for simplicity, i.e., ground truth source data is used as input. Our Orderless \textsc{Nade} training procedure is shown at the top of Figure~\ref{fig:diagram}.

\section{Block Gibbs Sampling}
\label{sec:gibbs}

The addition of the extra conditioning information used by Orderless \textsc{Nade} training requires an inference procedure to match. The goal of the inference procedure is to improve initial estimates as sampling goes on, leveraging earlier estimates as conditioning information for the next step. 

To that end, we use a block Gibbs sampling procedure at inference time. Standard Gibbs sampling is a Markov chain Monte-Carlo (MCMC) algorithm for estimating a multivariate probability distribution, whereby at each step one dimension is held out and the remaining dimensions are used as conditioning. Blocked Gibbs sampling holds out a set of dimensions instead of just one, as in standard Gibbs sampling.
Following Eq.~\ref{eq:onade_dist}, the goal of the proposed system is to model the conditional distribution of all sets of randomly masked sources, $\mathbf{s}_{\lnot D}$, given a mix, $\mathbf{m}$, and a corresponding set of complementary unmasked sources, $\mathbf{s}_{D}$, like $p_\theta(\mathbf{s}_{\lnot D} | \mathbf{s}_{D}, \mathbf{m})$. Therefore we consider each complete source waveform a discrete maskable variable (albeit containing many continuous dimensions) during the sampling process.
During inference, the mixture is always provided, unmasked, throughout the sampling process. Source estimates from previous steps are used as inputs for the next step, enabling the model to bootstrap its initial estimates for better results.

During sampling, an annealing schedule is used such that at earlier steps input sources are independently masked with a higher probability than later steps. At each step, sources are randomly masked with some probability, $\rho$, that decreases linearly over the sampling process according to the annealing schedule defined by Yao et. al.~\cite{yao2014equivalence}.
At step 0, $\rho = 1.0$ and at final step $N$, $\rho = 0.0$. We set $\alpha=0.95$. In this paper we test various values for the number of steps, $N$.

\begin{figure}[t]
    \centering
    \includegraphics[width=\columnwidth]{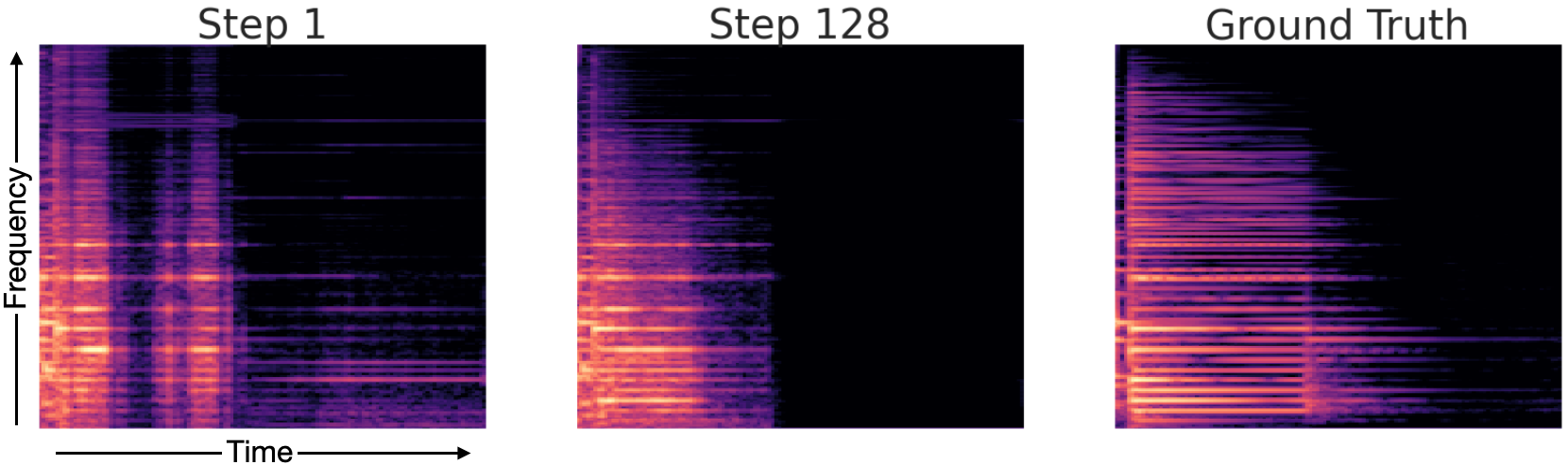}
    \caption{Spectrograms showing a piano estimate after the 1$^{\text{st}}$ Gibbs step (left) and the 128$^{\text{th}}$ Gibbs step (middle). A spectrogram showing the ground truth signal is on the right. The Gibbs sampling iteratively improves the source estimation.}
    \label{fig:spectrogram_example}
\end{figure}

\section{Experimental Validation}
\label{sec:experiments}

We conduct a set of experiments to validate the proposed training and testing setup. In our experiments, we test the effect of training a separation network using our Orderless \textsc{Nade} and block Gibbs sampling procedures.

In our main experiments, we vary the number of Gibbs sampling steps at inference time. These experiments are designed to give us an understanding of the dynamics of proposed block Gibbs sampling procedure for different lengths of sampling.
We measure the source separation performance after 1, 4, 16, 64, 128, 256, and 512 Gibbs steps. We train one network for each of the two datasets that we test: MUSDB18~\cite{musdb18} and Slakh2100~\cite{manilow2019cutting}.

The first dataset we examine is MUSDB18~\cite{musdb18}. MUSDB18 consists of 150 mixtures and corresponding source from real recording sessions featuring live musicians. 100 of these are for training, from which we reserve 10 as a validation set. The remaining 50 are used for evaluation. The second dataset we focus on is Slakh2100~\cite{manilow2019cutting}. Slakh2100 contains 2,100 mixtures with corresponding source data that were synthesized using professional-grade sample-based synthesis engines.
We train on 1289, use 270 for validation and evaluate on the 151 mixes in the test set.

\begin{figure*}[t]
    \centering
    \includegraphics[width=\textwidth]{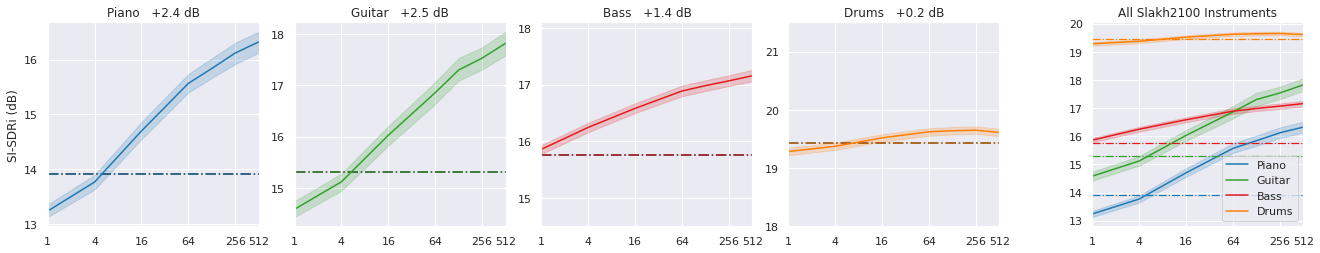}
    \includegraphics[width=\textwidth]{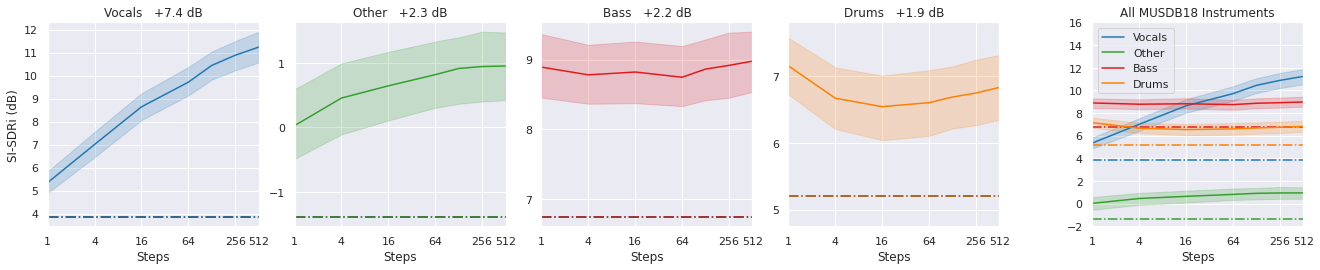}
    \caption{Separation performance, in terms of SI-SDR improvement (dB), for the Slakh2100 (top row) and MUSDB18 (bottom row) models after different numbers of Gibbs steps. Higher is better. Solid lines are means of our proposed system, shaded areas represent a 95\% CI, and dotted lines are mean of the baseline system. Note that the x-axis is log scaled. For many cases, the network is able to leverage earlier estimates to better performance. Furthermore, the sources that did not improve over time show large improvements over the baseline after just 1 step, indicating that Orderless \textsc{NADE} training was beneficial by itself.}
    \label{fig:gibbs_steps}
    \vspace{-0.5cm}
\end{figure*}

For both datasets, we downsample the audio to 16kHz. We segment the audio into 4 second windows with a 2 second hop. We only keep windows where 2 or more sources are active. For MUSDB18, we define a source as active if it has an RMS above --60dB, and all of the audio was converted to mono. We augmented MUSDB18 by applying pitch shifting and time stretching. For Slakh2100, we define an active source as having more than 5 note onsets in the corresponding MIDI. We did not use any augmentation with Slakh2100, and the audio is mono. We use the bass, drums, guitar, and piano sources from Slakh2100.

We implemented our own Demucs v2~\cite{defossez2019music} waveform-to-waveform architecture as our source separation system. Demucs follows a U-Net pattern, with 6 encoder and decoder layers that have skip connections to between corresponding encoder and decoder layers. At its bottleneck, Demucs has an bidirectional LSTM with the same dimensionality as the last encoder layer. Demucs outputs multiple sources as waveform data. We refer the reader to the Demucs paper for full details~\cite{defossez2019music}. In our implementation, we omitted many of the tricks that Demucs proposed to boost its performance (e.g., source remixing, weight rescaling, oversampling the audio). Demucs proposed a ``shift trick'', which averaged a fixed number of forward passes at random time offsets for inference, similar to our Gibbs sampling procedure. In our implementation, we found that the shift trick was detrimental to performance, and thus omitted it in our experiments. We trained using L1 loss on the waveform using Adam~\cite{kingma2014adam} with a learning rate of 3e-4 and batch size of 64 on 16 TPUv2 cores. We trained the Slakh2100 model for 100k steps and the MUSDB18 model for 85k steps.

Our Demucs was altered so that it had 8 additional input channels alongside the mixture. The first 4 channels were for injecting source estimates during training or sampling, and the final 4 channels (same dimensionality as the sources) provided as a sentinel flag to alert the network if any of the 4 input sources are masked. We compare this system to a baseline system trained in the typical manner, using only the mix as input, without Orderless \textsc{Nade} training or Gibbs sampling.

We conduct an additional experiment to test if models using the proposed training technique can effectively use the extra conditioning information. To do this we inject the ground truth source data for one source and measure the separation quality of the other sources after one Gibbs step. Because the model has perfect information for a given source, this serves as an upper bound on performance after one step. We
compare this to the case where all of the sources are masked at the first step. We perform this on the Slakh2100 dataset using the same model as the above Slakh2100 experiment.

We measure the performance of all of these systems using the improvement in scale-invariant source-to-distortion ratio (SI-SDRi)~\cite{le2019sdr} over the unprocessed mixture.

\begin{table}[]
\centering
\sisetup{table-format=1.1,round-mode=places,round-precision=1,table-number-alignment = center,detect-weight=true,detect-inline-weight=math,retain-explicit-plus=true}
\begin{tabular}{@{}lSSSS@{}}
\toprule
\multirow[b]{2}{1.375cm}{Injected GT Source} & \multicolumn{4}{c}{Estimated Source} \\   \cmidrule(l){2-5}
& {Piano}  & {Guitar}   & {Bass}   & {Drums}   \\ \midrule
                                     Piano  & {--}  & +7.08 & +4.73 & +2.77 \\ 
                                     Guitar & +6.22 & {--}  & +3.41 & +1.88 \\
                                     Bass   & +4.05 & +3.87 & {--}  & +2.38 \\
                                     Drums  & +1.77 & +2.09 & +2.31 & {--}  \\
\bottomrule
\end{tabular}
\caption{Increase over regular Gibbs sampling when injecting ground truth sources for the first Gibbs sampling step, on the Slakh test set, in terms of mean SI-SDRi (dB). Sources in the same pitch register, like piano and guitar, see the biggest increase.}
\label{tab:gt_injection}
\vspace{-0.5cm}
\end{table}

\section{Results and Discussion}
\label{sec:results}

The results of our main experiment are shown in Figure~\ref{fig:gibbs_steps}, which show plots with SI-SDR improvement as a function of number of Gibbs steps for each source in both datasets. The baseline system is shown as a dotted horizontal line for all sources. The top row shows results for the model trained on Slakh2100 sources and bottom row shows results for the model trained on MUSDB18 sources. The first thing that we note is that the proposed training and sampling procedure always beats the baseline model after the first 16 Gibbs steps for both datasets. In fact, across the two datasets we test, 6 of the 8 total sources show that more Gibbs steps produces better separation results that we test. However, our Gibbs sampling is not parallelized, so inference time scales linearly with the the number of Gibbs steps; after roughly 64 steps the sampling process has diminishing returns. Additionally, for MUSDB18, all sources show increased separation performance after the first Gibbs step--roughly +1--2 dB--indicating that the Orderless \textsc{Nade} training by itself is beneficial.

The most striking increase we see is from the vocals source in the MUSDB18 dataset, which increases by +7.4 dB after 512 Gibbs steps. The largest increases we see in the Slakh2100 dataset come from the piano and guitar sources, that improve by over +2.0 dB after 256 steps. These two sources can occupy the same pitch register in a mix, so we expect that information about one source will help when estimating the other. Drums in the Slakh2100 dataset is the only source that does not see significant improvement. We hypothesize that this is because the performance of the baseline Demucs model is already so high (it has the highest mean SI-SDRi across all sources in both datasets), so there is less room to improve.
Figure~\ref{fig:spectrogram_example} is an illustrative example showing spectrograms of a piano estimate after 1 Gibbs step and 128 Gibbs steps. 

The results in Figure~\ref{fig:gibbs_steps} indicate that a standard source separation architecture is able to bootstrap its way to better separation performance--even after training concludes--when the network is adapted to use our proposed training and inference procedure. This is an indication that network is able to learn and leverage dependencies between musical sources within a mixture.

In Table~\ref{tab:gt_injection}, we show the results of our experiment to determine how effective the training procedure is at producing a network that learns the conditional dependencies in a mix. Results after one Gibbs step show that injecting ground truth for any source improves the output of all other sources. This indicates that a model trained with our procedure has learned how to effectively use the extra input information. The largest boost comes when from estimating Guitar when ground truth Piano data is injected, and vice versa. These sources often have similar frequency content, meaning that any knowledge about one will frequently help the estimate of the other. 

\section{Conclusion}
\label{sec:conclusion}

In this work, our goal was to model the interdependent relationships between different sources in musical mixtures. This is desirable because most musical mixes contain sources that are intentionally highly coordinated. To this end, we applied an Orderless \textsc{Nade} training procedure and a block Gibbs sampling procedure for music source separation. An Orderless \textsc{Nade} setup enables us to conditionally model the relationship between any two sources in a musical mixture. The block Gibbs sampling procedure allows the network to boost the separation performance by bootstrapping from previous estimates. These training and sampling procedures are agnostic to network architectures, so we experimentally verified them using Demucs v2 on two source separation datasets. Our proposed training and sampling procedures increased performance for nearly all of the sources we tested. These results show the power of using even partial conditional information to analyze a musical scene. We are excited to see how other ideas from generative modeling can positively influence research in understanding musical scenes.

\bibliographystyle{IEEEbib}
\bibliography{references}

\end{document}